\documentclass[11pt,a4paper]{article}
\usepackage[utf8]{inputenc}
\usepackage{amsmath,amssymb}
\usepackage{graphicx}
\usepackage{geometry}
\usepackage{cite}
\usepackage{authblk}
\usepackage{xcolor}

\title{Controllable Chirality Sorting of Particles via Topological Optical Quasiparticles}
\author{Hao Zhang$^{1}$, Xi Xie$^{2,*}$, Yijie Shen$^{1,2,*}$}
\affil{$^{1}$Division of Physics and Applied Physics, School of Physical and Mathematical Sciences, Nanyang Technological University, Singapore, Singapore.\\$^{2}$Centre for Disruptive Photonic Technologies, The Photonics Institute, Nanyang Technological University, Singapore, Singapore.}
\date{\today}

\begin{document}

\maketitle

\begin{abstract}
The manipulation and sorting of chiral nanoparticles are of fundamental importance in multidisciplinary fields ranging from biochemistry to nanophotonics. In this study, we propose a novel and controllable chirality sorting mechanism for continuous particle separation using focused topological optical quasiparticles. Specifically, we investigate the sorting dynamics driven by tight-focused optical skyrmions and bimerons consisting of tailored spatial modes. By highly focusing free-space topological structure light fields, we generate intricate non-paraxial focal fields with tailored intensity and topological polarization textures. The sorting dynamics are systematically evaluated under the dipole approximation for fused silica nanoparticles. Our analytical calculation demonstrate that optical forces exert opposite directional pushes on particles of opposite chiralities ($\kappa = \pm 1$), enabling highly efficient spatial separation. Notably, we demonstrate that this sorting process is controllable; by tuning the topological charges, the sorting distance can be flexibly tailored and expanded. The dynamic sorting process in customized topological structures introduces a promising new paradigm for tunable, wide-range chirality sorting of micro- and nano-particles.
\end{abstract}

\section{Introduction}

The mechanical effects of light have fundamentally revolutionized the non-contact manipulation of micro- and nano-scale matter since the pioneering conceptualization of optical tweezers by Ashkin \cite{ashkin1970}. Over the decades, advances in the rigorous calculation and synthesis of optical forces have driven the rapid evolutionary progress of optical sorting techniques \cite{yang2025}, allowing researchers to continuously isolate target particles from bulk mixtures based on intrinsic physical traits. More recently, sophisticated structured light configurations and advanced material platforms have profoundly broadened the spectrum of advanced particle trapping strategies. Contemporary implementations now seamlessly utilize tailored vectorial fields \cite{wang2012, kritzinger2022, zhu2023}, cycloid-structured optical tweezers \cite{wei2023_ol}, and precision holographic optical surface-wave manipulations \cite{xie2022}. Furthermore, the frontier of nanoscale confinement has been pushed forward by advanced all-dielectric resonant nanotweezers \cite{dielectric_nanotweezers}. Accompanying these innovations are robust analyses of three-dimensional electromagnetic force controls \cite{optical_force_oe2025} alongside rigorous trajectory-based real-time force tracking methodologies \cite{dai2026_oe}, ensuring that adaptable optical manipulation operations remain a cornerstone of modern nanophotonics.

Beyond simple scalar properties, chirality is a pervasive fundamental signature in nature that specifies an entity's lack of mirror symmetry. The selective geometric recognition and spatial separation of chiral enantiomers are highly sought-after capabilities due to the drastically disparate behaviors of chiral isoforms observed in pharmacological responses and stereoselective catalysis. Currently, emerging optical configurations for sorting chiral particles broadly capitalize on highly customized chiral light fields \cite{shi2026}, seeking to exploit specialized optical chiral gradient forces formulated by the subtle localized exchange of optical spin densities \cite{cao2018}. For example, recent techniques have successfully achieved the switchable rotation of metallic nanostructures utilizing intensity chirality-invariant focus fields \cite{chen2023_ol} and demonstrated all-optically controlled holographic plasmonic vortex arrays for multiplexed metallic particle manipulations \cite{ju2023_ol}. Even so, a persistent challenge resides in crafting structurally highly tunable, dynamically robust, and macroscopically controllable optical sorting fields capable of achieving long-range, continuous spatial isolation for diverse chiral samples.

In parallel with nanomechanical advancements, the realization of topological quasiparticles—such as skyrmions and bimerons—has stimulated remarkable explorations in both fundamental and applied photonics. While topological defect lattices have recently demonstrated unprecedented capabilities for localized on-chip trapping and sorting within magnetic architectures \cite{xu2025}, their photonic analogues hold equally transformative potential. Originating conceptually from high-energy physics, these optical textures map neatly into the electromagnetic realm via the spatial distributions of full-Poincaré Stokes vectors \cite{shen2024, gao2020}. These inherently non-trivial free-space structured light fields contain intricate, customized networks of polarization mappings and phase singularities capable of mediating distinct, multidimensional spin-orbit force interactions \cite{odonnell2024}. Consequently, the exotic analytical forces and torques derived from localized spinning light \cite{canaguier2013} and tightly confined skyrmionic geometries suggest the capacity for breakthrough mechanical separations. Yet, their direct utility in the spatially continuous optical chiral sorting of specific enantiomer populations remains relatively unexploited.

In this work, we propose the application of topological optical quasiparticles—specifically, high-numerical-aperture focused optical skyrmions and bimerons—for the spatially controllable sorting of chiral nanoparticles. By configuring orthogonal superposition states natively derived from Laguerre-Gaussian (LG) states, we quantitatively analyze the chiral separation phenomena. We identify disparate, enantiomer-specific optical gradient forces that dispatch particles along disparate trajectories. Crucially, we demonstrate that scaling the collective topological charge of the excitation field actively controls the localized optical chirality density, which actively governs the net particle sorting gap, culminating in a functionally generalized parameter space for structured-light-driven chirality sensing and continuous separations.

\section{Theoretical Framework and Methods}

\subsection{Free-Space Topological Optical Textures}
To construct the tailored topological textures mathematically, we initially superimpose two paraxial structured light beams possessing orthogonal polarizations \cite{gao2020}. Specifically, the input generalized paraxial field state $|\psi\rangle$ forming an optical skyrmion is modeled as the coherent vectorial summation of Laguerre-Gaussian (LG) eigenmodes, defined by:
\begin{equation}
    |\psi\rangle = |LG_{p_1,\ell_1}\rangle |R\rangle + |LG_{p_2,\ell_2}\rangle |L\rangle,
    \label{eq:superposition}
\end{equation}
where $|R\rangle$ and $|L\rangle$ stand for the right- and left-handed circular polarization basis states, respectively. Here, $p_{1,2}$ formulate the radial indices while $\ell_{1,2}$ label the topological charges (azimuthal indices). The explicit expression for individual constituent LG cylindrical modes in coordinates $(\rho, \theta, z)$ takes the general form:
\begin{equation}
    \begin{aligned}
    LG_{p,\ell} (\rho, \theta, z) &= \sqrt{\frac{2p!}{\pi (p+|\ell|)!}} \frac{1}{w(z)} \left[ \frac{\sqrt{2}\rho}{w(z)} \right]^{|\ell|} \text{e}^{-\frac{\rho^2}{w^2(z)}} \text{e}^{i\ell\theta} \\
    &\quad \times L_p^{|\ell|} \left[ \frac{2\rho^2}{w^2(z)} \right] \text{e}^{ikz - i(2p+|\ell|+1)\vartheta(z)},
    \end{aligned}
    \label{eq:lg_beam}
\end{equation}
where $w(z)$ encapsulates the beam waist scaling at the propagation plane $z$, $L_p^{|\ell|}$ signifies the generalized Laguerre polynomial, $k=2\pi/\lambda$ represents the wavenumber for wavelength $\lambda$, and $\vartheta(z)$ delineates the Gouy phase shift. Alternatively, the bimeron states incorporate modes carrying mutually orthogonal linear polarizations, specifically $|H\rangle$ (horizontal) and $|V\rangle$ (vertical), maintaining a structurally comparable synthesis mechanism \cite{shen_tutorial}. 

The spatial polarimetric mappings of these fields translate directly into topological optical quasiparticles by computing the normalized Stokes parameters. The continuous evolution of the localized spatial polarization is graphically embedded utilizing the normalized spatial Stokes vector defined by 
\begin{equation}
    (\mathcal{S}_x, \mathcal{S}_y, \mathcal{S}_z) = (\mathcal{S}_1/\mathcal{S}_0, \mathcal{S}_2/\mathcal{S}_0, \mathcal{S}_3/\mathcal{S}_0),
    \label{eq:stokes}
\end{equation}
where the fundamental Stokes scalars ($\mathcal{S}_0, \mathcal{S}_1, \mathcal{S}_2, \mathcal{S}_3$) denote the local field intensity ($\mathcal{S}_0$), along with projections onto orthogonal linear ($\mathcal{S}_1, \mathcal{S}_2$) and circular ($\mathcal{S}_3$) canonical bases \cite{shen_tutorial}.

\subsection{Tight Focusing via Richards-Wolf Integral}
To obtain the extremely localized intense gradient fields strictly required for nanoscale manipulations, the paraxial coherent combinations formally traverse an aplanatic objective lens conforming to a high numerical aperture (here, $\text{NA} = 0.9$). Operating physically in the rigorous focal regime, the exact electromagnetic spatial variations are obtained by executing the strict vectorial Debye-Wolf framework \cite{chen2025}. The sharply localized focused electric field $\mathbf{E}_f$ defined at the cylindrical destination coordinates $(r_f, \phi, z_f)$ is formulated by the well-known Richards-Wolf integral \cite{chen2025}:
\begin{equation}
    \mathbf{E}_f (r_f, \phi, z_f) = \frac{i}{\lambda} \int_{0}^{\theta_{\text{max}}} \int_{0}^{2\pi} \mathbf{E}_{\Omega}(\theta, \varphi) \cdot \exp [ikr_f \sin \theta \cos (\varphi - \phi) + ikz_f \cos \theta] \sin \theta \, d\theta \, d\varphi,
    \label{eq:rw_integral}
\end{equation}
where $\mathbf{E}_{\Omega}(\theta, \varphi)$ projects the field amplitudes parameterized over the objective reference sphere defined by the geometric polar ($\theta$) and azimuthal ($\varphi$) coordinates. The maximum semi-aperture angle $\theta_{\max}$ inherently relates to the optical numerical aperture directly via $\sin\theta_{\max} = \text{NA}/n$. The companion rigorous focused magnetic field $\mathbf{H}_f$ is calculated by integrating the matching curl relation identically. We establish the simulation constraints surrounding an external liquid index equivalent to water ($n \approx 1.33$) while probing dynamics exclusively at incident optical wavelength $\lambda = 532$ nm.

\subsection{Time-Averaged Optical Forces on Chiral Enantiomers}
Within the localized limit whereby the particle's spherical radius ($R=40$ nm) falls drastically beneath the operational wavelength ($2R \ll \lambda$), the resultant optical momentum transfers can be rigorously uncoupled via the dipolar formulation \cite{man2024}. Enantiomeric nano-entities map classically into polarizable equivalent point dipoles equipped consistently with mutually coupled electromagnetic chiral cross-terms. The comprehensive expression for the time-averaged total optical force $\langle \mathbf{F} \rangle$ experienced by the specific chiral dipole equates to \cite{canaguier2013}:
\begin{equation}
    \begin{aligned}
    \langle \mathbf{F} \rangle &= \nabla U + \frac{\sigma n}{c} \langle \mathbf{S} \rangle - \text{Im}(\chi) \nabla \times \langle \mathbf{S} \rangle - \frac{c\sigma_e}{n} \nabla \times \langle \mathbf{L}_e \rangle \\
    &\quad - \frac{c\sigma_m}{n} \nabla \times \langle \mathbf{L}_m \rangle + \omega \gamma_e \langle \mathbf{L}_e \rangle + \omega \gamma_m \langle \mathbf{L}_m \rangle \\
    &\quad + \frac{c k^4}{12 \pi n} \text{Im}(\alpha_e \alpha_m^*) \text{Im}(\mathbf{E} \times \mathbf{H}^*).
    \end{aligned}
    \label{eq:force}
\end{equation}
Across Equation \eqref{eq:force}, individual terms relate to discrete force regimes mapping structurally to the input topological optical architecture. The constituent parameter dependencies explicitly include the local optical scalar potential ($U$), the classic time-averaged optical Poynting vector dictating scattering momentum transitions ($\langle \mathbf{S} \rangle$), distinct time-averaged electric and magnetic orbital/spin angular momentum spatial densities ($\langle \mathbf{L}_e \rangle$ and $\langle \mathbf{L}_m \rangle$) driving complex vortical torques, and finally, the local tightly focused complex electromagnetic harmonic vectorial profiles ($\mathbf{E}$ equating to $\mathbf{E}_f$, alongside corresponding $\mathbf{H}$). Regarding the explicit chiral specimen specifications parameterized universally across the equations: $c$ tracks light's free-space velocity, $n$ acts as the surrounding suspension's bulk structural index, and $\omega$ specifies the underlying temporal angular frequency. Extracted particle-specific variables include the composite absorption cross-section $\sigma$, along with decoupled electric ($\sigma_e$) and magnetic ($\sigma_m$) specific dispersion cross-sections alongside electric and magnetic polarizabilities ($\alpha_e$ and $\alpha_m$, respectively). Specifically, the intrinsic degree of molecular chirality directly associates the governing factors ($\chi, \gamma_e, \gamma_m$) mapping directly phenomenologically in subsequent tracking models by distinguishing mirror-imaged particles via isolated chiral parameters denoted cleanly by opposing polarities ($\kappa=\pm1$). 

We concentrate our computational scope towards isolated fused silica nano-enantiomers exhibiting geometric spherical radius equivalent to $40$ nm coupled against a standard constituent bulk relative index $n=1.4607$. Complete continuous macroscopic spatial tracking operations leverage numeric finite-element multiphysics packages natively (COMSOL Multiphysics), operating cleanly within viscous aqueous regimes. The resulting data isolates analysis towards the pure lateral transversal dynamics—specifying $F_x, F_y$ plane components sequentially—strategically ignoring the uncoupled longitudinal $F_z$ axis to rigorously outline the spatially tunable two-dimensional continuous sorting footprint natively constrained within the localized two-dimensional plane.

\section{Results and Discussion}
\begin{figure}[t]
    \centering
    \includegraphics[width=1\linewidth]{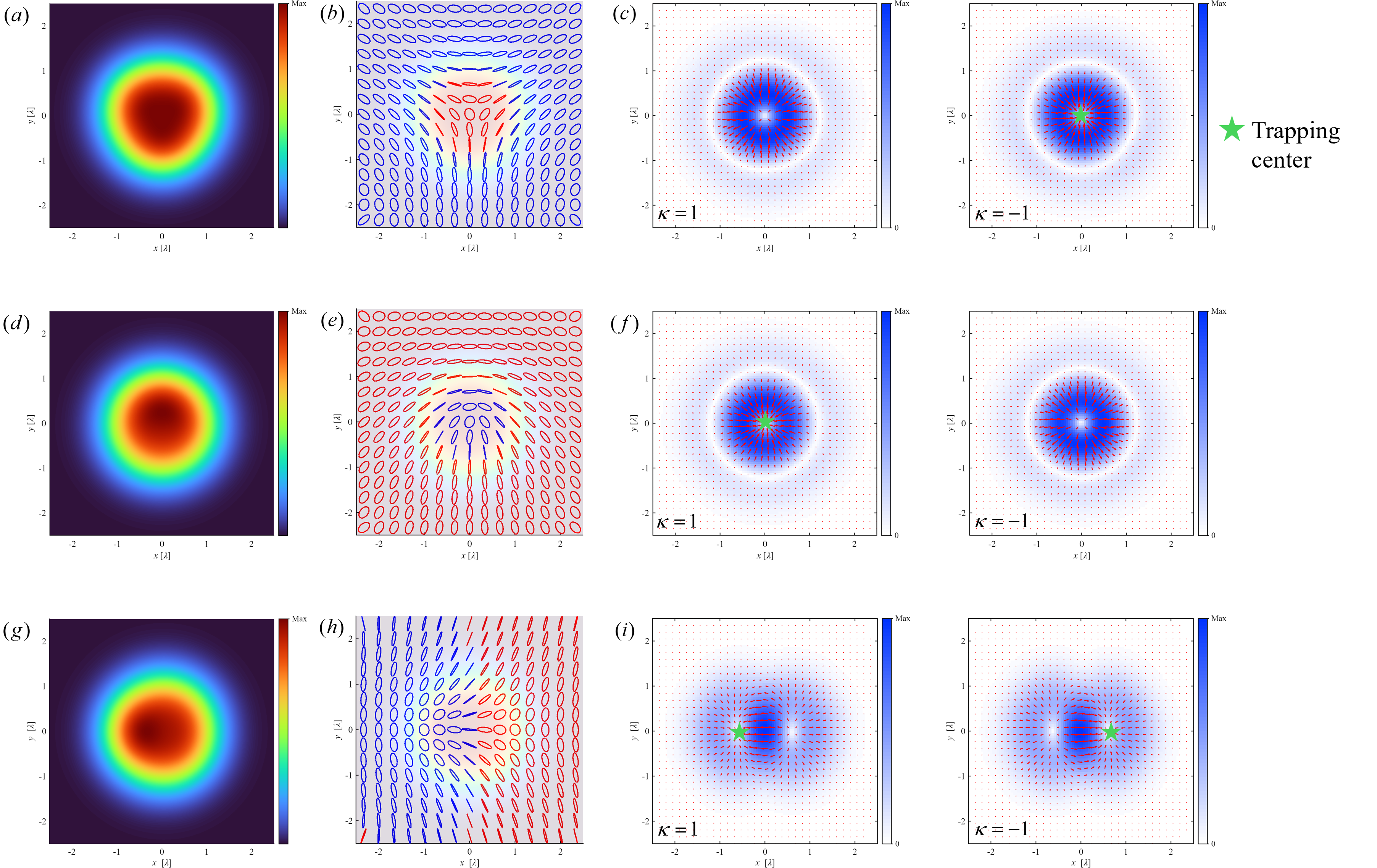}
    \caption{The intensity, polarzation and optical force profile after tight focusing. The first row corresponds to $LG_{0,0}|R\rangle+LG_{0,1}|L\rangle$, the second row corresponds to $LG_{0,1}|R\rangle+LG_{0,0}|L\rangle$ and the third row corresponds to $LG_{0,0}|H\rangle+LG_{0,1}|V\rangle$. Right circles represent the right handedness of polarization while the blue circles represent left handedness. The green star indicates the trapping center.}
    \label{fig:placeholder}
\end{figure}
\subsection{Topological Profiles and Chiral Sorting Fields}

The generation of highly confined spatial topological features enables the customized directional routing of chiral particles. Figure 1 elaborates the focal plane spatial characteristics of optical skyrmions and bimerons alongside the ensuing lateral optical force maps. The figures present the intensity distributions (Fig. 1 (a), (d), (g)) and spatial polarization topographies (Fig. 1 (b), (e), (h)) of the tightly focused fields. The respective optical force vectors ($F_x$ and $F_y$) mapping to differently handed enantiomers ($\kappa = \pm 1$) are illustrated by red arrows in the remaining panels.

We investigated three primary topological configurations. In the first configuration (Fig. 1, 1st Row), an optical skyrmion is formed by synthesizing a left-handed (L) circularly polarized optical vortex ($LG_{0,1}$) and a right-handed (R) polarized Gaussian beam ($LG_{0,0}$). The optical force maps dictate that particles with $\kappa=1$ and $\kappa=-1$ experience oppositely directed chiral lateral components within Eq. \eqref{eq:force}, forcing them towards divergent orbital pathways that establish the localized foundation for separation. The second configuration (Fig. 1, 2nd Row) inverts the modes into a left-handed Gaussian ($LG_{0,0}$) and a right-handed vortex ($LG_{0,1}$) beam. By exchanging the polarization states of the central and peripheral parts of the smectic molecules, particles with opposite handedness (with $\kappa=1$) can be captured, and particles with $\kappa=-1$ can be dispersed. Two types of skyrmions respectively produced corresponding chiral trapping centers.

Conversely, an optical bimeron (Fig. 1, 3rd Row) is structured using two orthogonal linear polarization states: an incident horizontally polarized ($|H\rangle$) Gaussian beam ($LG_{0,0}$) superimposed with a vertically polarized ($|V\rangle$) optical vortex ($LG_{0,1}$). The tight focusing under high NA yields an asymmetric dual-lobed intensity structure globally distinct from skyrmion profile. The spatially varying lateral forces homogeneously sweep the targeted $\kappa=1$ particles towards one external high-intensity , critically dragging the remaining respective $\kappa=-1$ particles progressively toward the another trapping center. This unique consequence constructs an active framework for continuous side-by-side spatial stereoisomer discrimination.

\subsection{Dynamic Sorting in Skyrmions and Bimerons}

To validate continuous practical sorting capacities functionally and dynamically, we natively modeled corresponding longitudinal translational tracking frameworks charting enantiomer paths continuously against prevailing solvent viscous forces structurally via COMSOL.  Figure 2 delineates tracked sequential dynamic separations within pure skyrmionic and bimeronic quasi-particle architectures across progressive temporal increments.

\begin{figure}[h]
    \centering
    \includegraphics[width=1\linewidth]{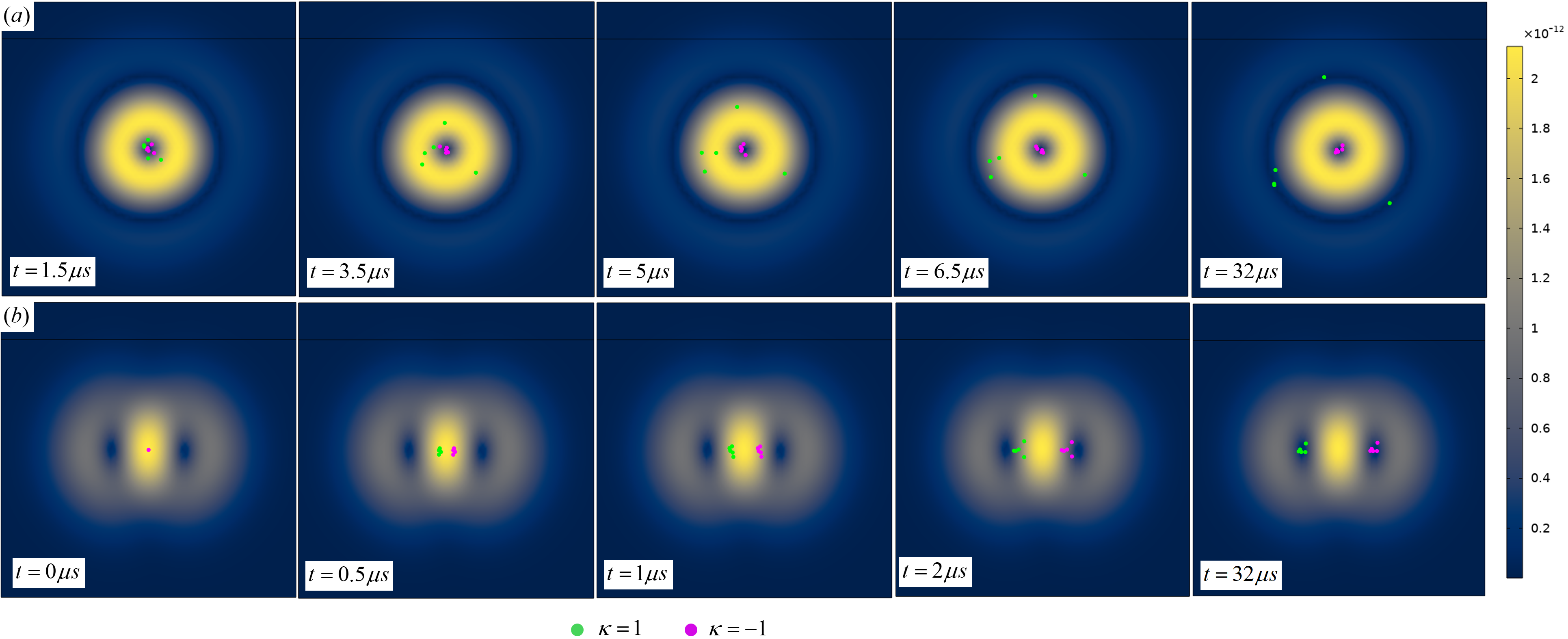}
    \caption{Simulation of particles' motion. The first row corresponds to the motion driven by skyrmion ($LG_{0,0}|R\rangle+LG_{0,1}|L\rangle$), the second row corresponds to bimeron ($LG_{0,0}|H\rangle+LG_{0,1}|V\rangle$) results. Green and purple dots represent particles with opposite chirality ($\kappa=1$ and $\kappa=-1$), respectively.}
    \label{fig:placeholder}
\end{figure}

To accurately model the continuous tracking and chiral sorting processes dynamically within the viscous fluid, we employed the Langevin equation of motion. The governing dynamic equation for the nanoparticle is given by:
\begin{equation}
    m \frac{d\mathbf{v}}{dt} + 6\pi\eta a(\mathbf{v} - \mathbf{u}) = \mathbf{F}_{\text{opt}} + \mathbf{F}_{\text{Brownian}},
    \label{eq:langevin}
\end{equation}
where $m$ is the mass of the chiral nanoparticle, $\mathbf{v}$ denotes its translational velocity vector, and $t$ defines the time scalar, $\eta$ identifies the dynamic viscosity of the surrounding water medium, $a$ refers to the spherical radius of the particle (equivalent to $R = 40$ nm), and $\mathbf{u}$ specifies the velocity field of the bulk fluid (taken as $\mathbf{u} = 0$ for a stationary aqueous environment). On the right-hand side, $\mathbf{F}_{\text{opt}}$ embodies the highly tailored time-averaged optical sorting force (derived analytically as $\langle \mathbf{F} \rangle$ in Eq. 5), accelerating the targeted enantiomers along selective topological pathways. Finally, $\mathbf{F}_{\text{Brownian}}$ represents the stochastic Brownian force accounting for thermal agitation from surrounding solvent molecules, inducing random diffusive motions across the nanoparticle tracking trajectories.

\subsection{Controllable Sorting via Higher-Order Topology Charges}
\begin{figure}[t]
    \centering
    \includegraphics[width=0.9\linewidth]{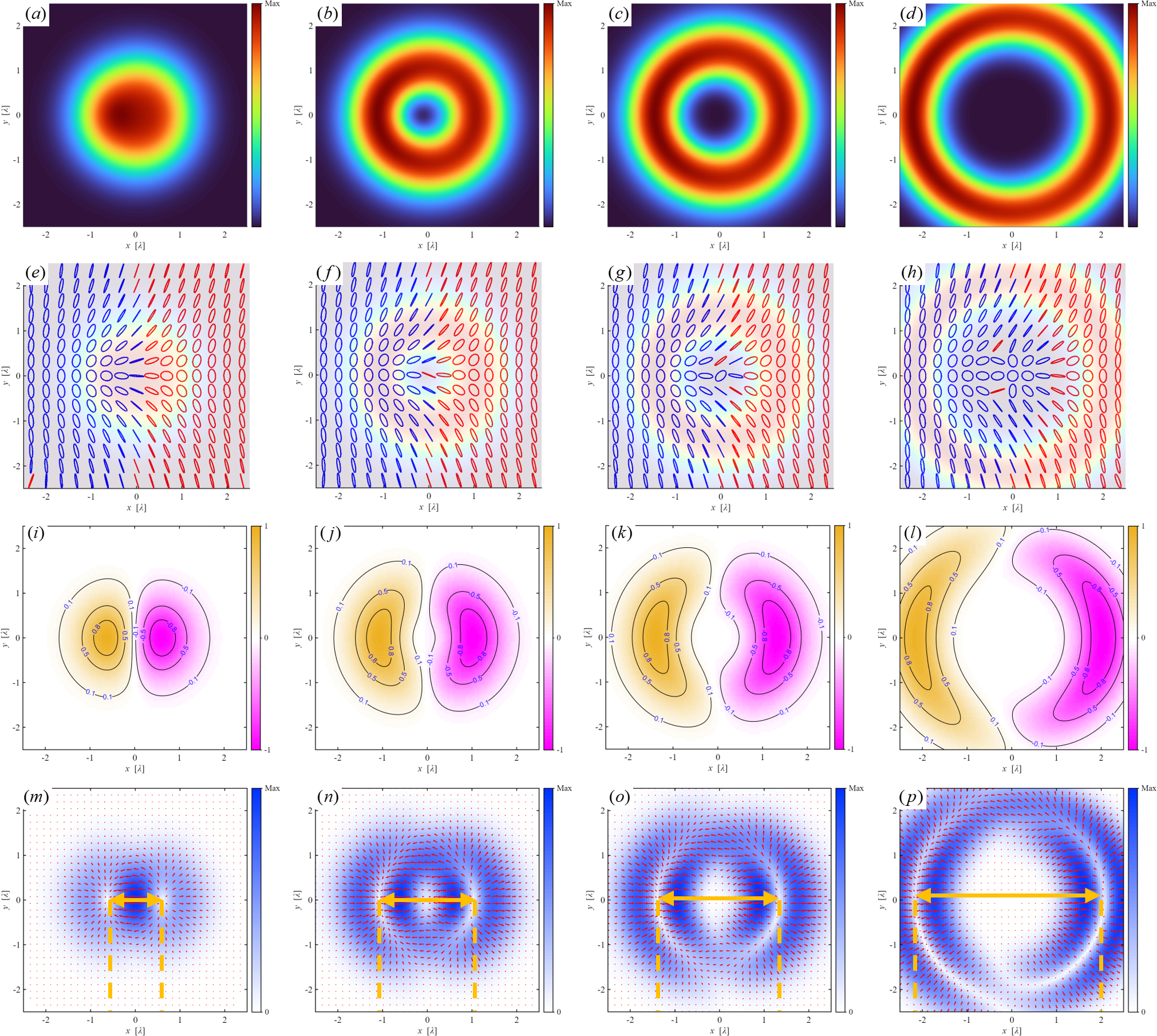}
    \caption{Different sorting profile by manipulating the the topology charges. Based on bimeron sorting profile (the first column), the second to the fourth columns corresponds to $LG_{01}|H\rangle+LG_{02}|V\rangle$, $LG_{02}|H\rangle+LG_{03}|V\rangle$ and $LG_{05}|H\rangle+LG_{06}|V\rangle$, respectively.}
    \label{fig:placeholder}
\end{figure}
A key analytical capability exposed by our topological tuning paradigm involves managing deterministic spatial separation distance efficiently by  modulating simple initial field input terms directly. Furthermore, the underlying physical mechanism driving this enantiomeric segregation can be comprehended strictly by evaluating the localized chiral components of the optical potential $U$, which dictates the primary optical gradient force ($\mathbf{F}_{\text{grad}} = \nabla U$). The complete scalar optical potential field within the dipole limit is expressed as:
\begin{equation}
    U = \frac{1}{4}\text{Re}(\alpha_e)|\mathbf{E}|^2 + \frac{1}{4}\text{Re}(\alpha_m)|\mathbf{H}|^2 + \frac{1}{2}\text{Re}(\chi)\text{Im}(\mathbf{E} \cdot \mathbf{H}^*).
    \label{eq:opt_potential}
\end{equation}
The first two terms identify the achiral electric and magnetic interactions corresponding respectively to the local field intensity topographies. Crucially, the third localized term formulates the chiral optical potential, selectively orchestrating the enantiomer-specific sorting effects. The spatial distribution of this chiral potential relies heavily on the generated optical chirality density $C$, mathematically defined as:
\begin{equation}
    C = -\frac{\omega\varepsilon_0}{2}\text{Im}(\mathbf{E}^* \cdot \mathbf{B}),
    \label{eq:chirality_density}
\end{equation}
where $\varepsilon_0$ denotes the vacuum permittivity and $\mathbf{B}$ identifies the magnetic flux density. Utilizing the conjugate symmetry relationship $\text{Im}(\mathbf{E}^* \cdot \mathbf{H}) = -\text{Im}(\mathbf{E} \cdot \mathbf{H}^*)$, it becomes evident that the third term of the optical potential in Eq. \ref{eq:opt_potential} is strictly proportional to the optical chirality density $C$ (scaling proportionally as $\sim \frac{c^2}{\omega} \text{Re}(\chi) C$, where $c$ is the speed of light). The real part of the chiral polarizability, $\text{Re}(\chi)$, adopts explicitly opposite signs for opposite enantiomers ($\kappa = \pm 1$). Consequently, the gradient of the tailored optical chirality density ($\nabla C$) drives the alternative chiral species cleanly toward completely divergent steady-state potential wells.

Figure 3 systematically depicts this controllable sorting mechanism scaling functionally through consecutively augmenting the constituent azimuthal indices forming the topological beams. Starting from the baseline bimeronic configuration in Column 1, we map sequential topological extensions extending ultimately up to structural combinations of $LG_{0,5}$ and $LG_{0,6}$ presented in Column 4. Notably, the simulated arrays in Figures 3(i--l) meticulously reveal the focal plane distributions of the corresponding optical chirality density $C$.

With the persistent increment of the targeted topological charges, the overall optical field expands radially, forming significantly larger topological profiles. Consequently, the optical chirality density $C$ spreads outward (as evidenced sequentially transitioning through Fig. 3i to 3l), smoothly relocating the corresponding gradient maxima and minima regions natively to much wider focal coordinates. Because the chiral optical potential wells track directly against the spatial spread of $C$, the steady-state equilibrium trapping targets equivalently scale symmetrically outward. This guarantees a direct, quantifiable correlation where structural macroscopic expansion translates seamlessly into broader physical sorting distances separating the individual enantiomers. Such predictable manipulation of $C$ allows experimentalists to dynamically magnify the lateral chiral sorting boundaries identically correlating with the incident topological index, eliminating cumbersome hardware replacements completely.

\begin{figure}[h]
    \centering
    \includegraphics[width=1\linewidth]{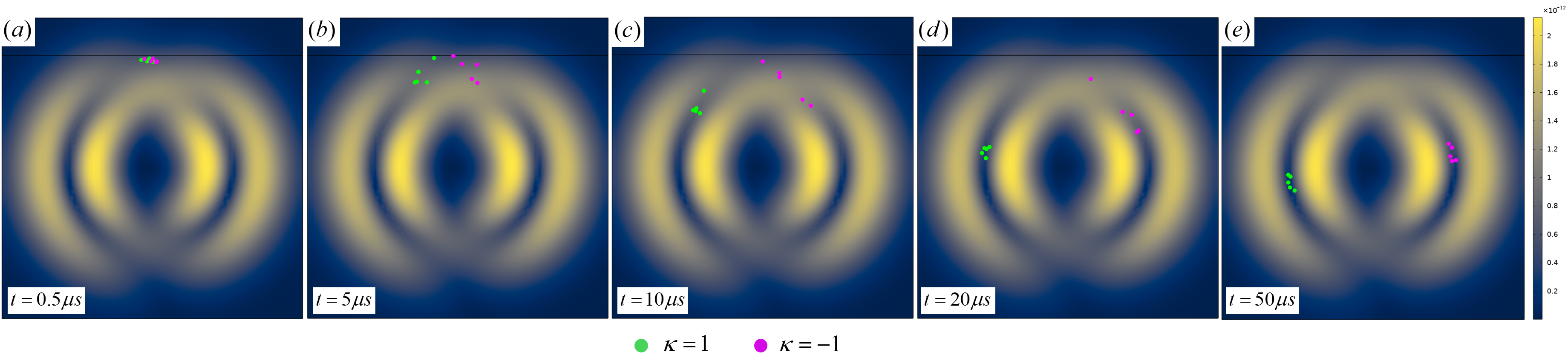}
    \caption{Simulation of high order trapping profile.}
    \label{fig:placeholder}
\end{figure}
Finally, explicitly demonstrating corresponding continuous spatial kinematic behavior , Figure 4 fully explicitly validates dynamic capture through targeted high-order configurations  ($LG_{0,2}|H\rangle$ and  $LG_{0,3}|V\rangle$ combination). Because the force field intensity of this high-order topological configuration center is too weak, it is not conducive to the actual chiral separation process. Therefore, in the simulation, the situation where chiral particles enter the light field from the outside was considered.

\section{Conclusion}

In summary, we have proposed and numerically demonstrated a controllable, highly tunable chirality sorting technique utilizing accurately engineered optical topological quasiparticles. Through the integration of highly focused optical skyrmions and bimerons synthesized from elementary LG spatial modes, robust and localized lateral optical force fields are generated which elicit remarkably diverse enantiomer-specific kinetic responses under the rigorous dipole framework. Our dynamic COMSOL tracking simulations verify that 40-nm fused silica nanoparticles representing opposing chiralities ($\kappa = \pm 1$) can be completely and continuously separated into distinct steady-state localized spatial coordinates laterally. Most importantly, we demonstrated that the absolute macroscopic spatial sorting capacity—the absolute transverse distance permanently separating the purified enantiomers—can be seamlessly configured and expanded organically by incrementally raising the combined topological charges natively embedded inside the incident optical field configurations. This uniquely adaptable topological technique establishes an active generalized optical toolbox for dynamic non-destructive nanoscale manipulations, offering highly promising configurations supporting integrated next-generation high-throughput enantiomeric screening arrays and customized continuous lab-on-a-chip biosensing implementations.

\end{document}